\documentclass{article}
\usepackage{spconf,amsmath,graphicx}
\usepackage{amssymb}
\usepackage{multirow}
\usepackage{booktabs}

\title{DEEP CROSS-MODAL STEGANOGRAPHY USING NEURAL REPRESENTATIONS}

\name{Gyojin Han, Dong-Jae Lee, Jiwan Hur, Jaehyun Choi, and Junmo Kim}
\address{School of Electrical Engineering, KAIST, South Korea}

\begin{document}
\maketitle
\begin{abstract}
Steganography is the process of embedding secret data into another message or data, in such a way that it is not easily noticeable. With the advancement of deep learning, Deep Neural Networks (DNNs) have recently been utilized in steganography. However, existing deep steganography techniques are limited in scope, as they focus on specific data types and are not effective for cross-modal steganography. Therefore, We propose a deep cross-modal steganography framework using Implicit Neural Representations (INRs) to hide secret data of various formats in cover images. The proposed framework employs INRs to represent the secret data, which can handle data of various modalities and resolutions. Experiments on various secret datasets of diverse types demonstrate that the proposed approach is expandable and capable of accommodating different modalities.
\end{abstract}
\section{Introduction}
\label{sec:intro}
Steganography is a technique that involves embedding secret data, such as binary messages, images, audio, or video, into another message or data in a way that makes it difficult to detect. The objective of steganography is to add an extra layer of security to communication, making it harder for unauthorized parties to detect the presence of hidden information.

Recently, there has been a surge of interest in utilizing deep neural networks (DNNs) in steganography pipelines. Many of these approaches, such as those outlined in \cite{NIPS2017_838e8afb, Zhu_2018_ECCV, NEURIPS2020_73d02e43, Lu_2021_CVPR, 9676416}, use DNNs as encoders and decoders for embedding and extracting hidden information. These deep steganography techniques, which involve end-to-end training of DNNs, have demonstrated advantages over traditional steganography methods in terms of capacity and security, while also being simpler to implement.

\begin{figure}[t]
\begin{center}
\centerline{
\includegraphics[width=1.0\columnwidth]{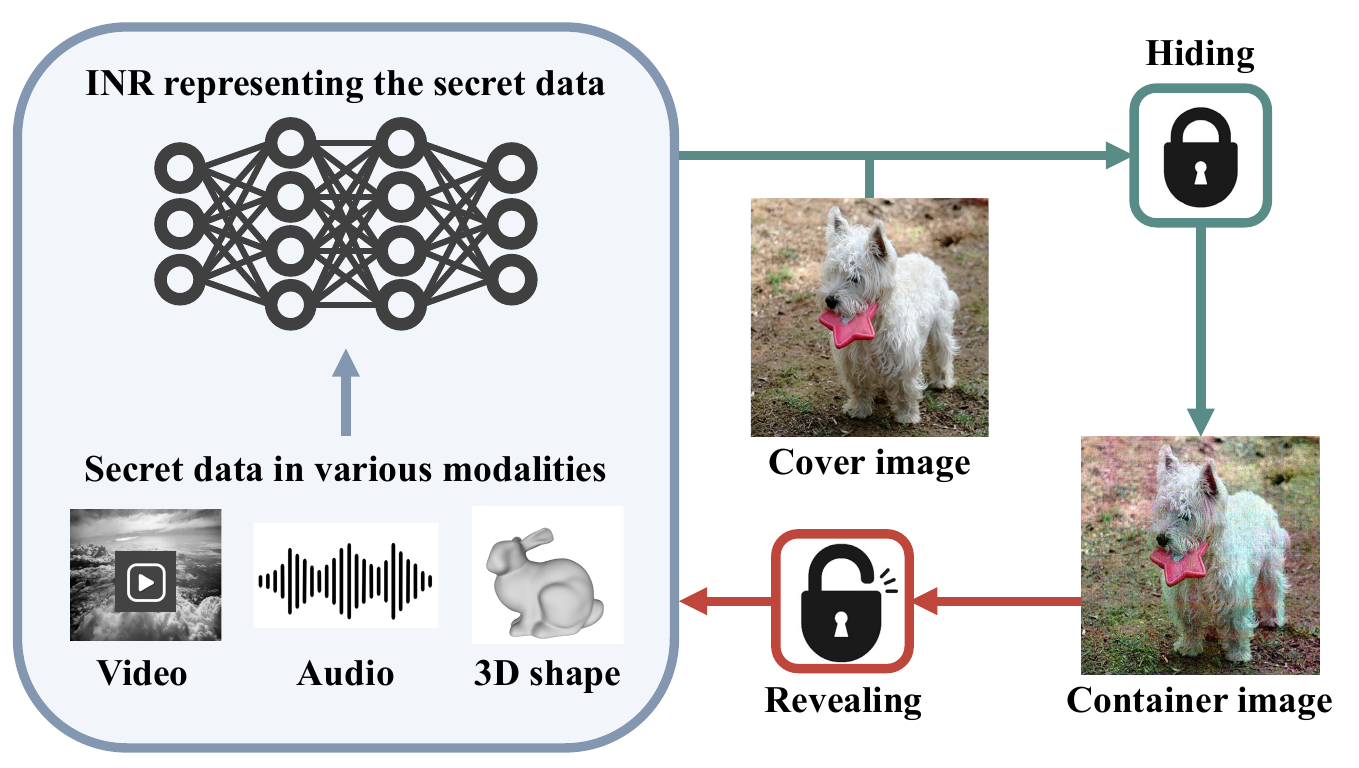}
}
\caption{We provide a comprehensive solution for deep cross-modal steganography dealing with secret data of various modalities by using INRs.}
\label{concept}
\end{center}
\end{figure}

\begin{figure*}[ht]
\begin{center}
\centerline{
\includegraphics[width=\linewidth]{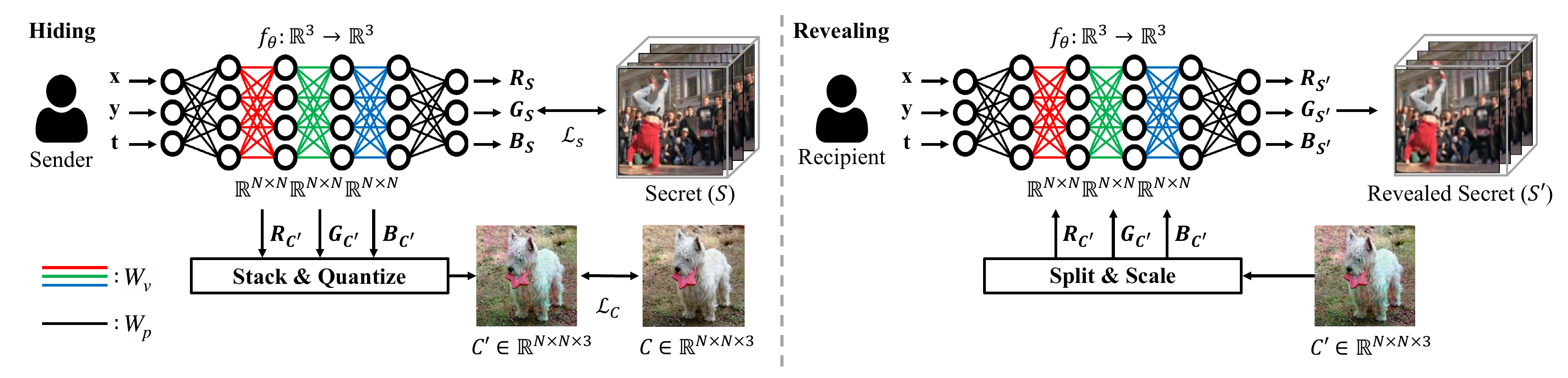}
}
\caption{We train $f_\theta$ to approximate secret data $S$ with various modalities, with a constraint that a portion of weights $\mathbf{W}_v$ to imitate the cover image $\mathbf{C}$. Using a container image $\mathbf{C}'$, the only recipient who knows the remaining portion of the parameters $\mathbf{W}_p$ can reveal the secret data $\mathbf{S}'$ by reconstructing the whole INR network.}
\label{method}
\end{center}
\end{figure*}

However, most of these methods are limited in scope and focus only on specific data types, so they do not provide a comprehensive solution for data hiding. The majority of deep steganography approaches concentrate on hiding binary messages \cite{NIPS2017_fe2d0103, Zhu_2018_ECCV, Tancik_2020_CVPR} or natural images \cite{NIPS2017_838e8afb, NEURIPS2020_73d02e43}. While some efforts have been made to use deep steganography to hide video \cite{10.1145/3323873.3325011, mishra2019vstegnet} or audio \cite{kreuk2019hide} within the same data type, there is a shortage of research on cross-modal steganography. This type of steganography involves hiding secret data in cover data that is in a different format than the secret data. This can be particularly challenging when the secret data has higher dimensions than the cover data, such as hiding a video in an image.

To address these limitations, we establish a new deep cross-modal steganography framework using Implicit Neural Representations (INRs) that can be generalized to tasks with secret data of various formats. We employed INRs to represent the secret data, as INRs have the ability to handle data of various modalities and resolutions as seen in Fig.~\ref{concept}.
Under this framework, the objective is to send secret data expressed in INR through a container image with minimal distortion from a cover image.
To achieve this goal, we make the following contributions:
\begin{itemize}
\item We enable the sender and recipient to share a base network and transmit a portion of the weights constituting that network in the form of an image. 
\item To reduce quantization error that occurs during the conversion process from the weights to the images, we propose a simplified quantization-aware training.
\end{itemize}
We demonstrate the expandability of our method and potential for various applications through experiments on datasets of various modalities.

\section{Method}
\label{sec:method}
\subsection{Overview}
Our goal is to solve the cross-modal steganography problem, specifically hiding secret data $\mathbf{S}$ of various modalities (such as a video, audio, or 3D shape) into a single cover image $\mathbf{C}$.
The container image $\mathbf{C'}$ including the information about $\mathbf{S}$ should be difficult to distinguish from the cover image $\mathbf{C}$ by human observation. In our framework, the sender and recipient share the architecture of an INR, referred to as the base network. The base network is composed of variable weights ($\mathbf{W}_v$) and pre-defined weights ($\mathbf{W}_p$). The hiding stage is carried out by training the INR so that the container image $\mathbf{C'}$ converted from the $\mathbf{W}_v$ is close to the cover image $\mathbf{C}$, while simultaneously making the revealed secret $S'$ reconstructed by the INR to be close to the secret $S$. 
The decoding can be accomplished by reconstructing the entire INR using $\mathbf{W}_v$ which is converted from $\mathbf{C}'$ and $\mathbf{W}_p$ which is pre-defined.
This is achieved by treating each channel of the container image as a weight matrix and inserting the weight matrices with converted image channels in the base network.

\subsection{Deep Cross-Modal Hiding Framework}
In this paper, we consider, but not limited to, a multi-layer perceptron (MLP) with $L$ layers as an example of INR $f_\theta$ where the parameter $\theta$ can be defined as a set of matrices such that $\theta = \{\mathbf{W}_l|\mathbf{W}_l \in \mathbb{R}^{C_{in} \times C_{out}}\}_{l=0}^{L-1}$ and $C_{in}$ and $C_{out}$ represent the number of units in each layer before and after $\mathbf{W}_l$. 
For the proposed method, the sender and recipient share the pre-defined weights $\mathbf{W}_p$ and positions of variable weights. Formally, the variable weights $\mathbf{W}_v$ refer to a subset of the weights constituting the base network that can be trained. Additionally, they share the hyperparameters $w_{min}$ and $w_{max}$, which are required for scaling and quantization.

We create a container image $\mathbf{C'}$ by stacking the $\mathbf{W}_v$
% , $\mathbf{W}_{l_1}$, $\mathbf{W}_{l_2}$, and $\mathbf{W}_{l_3}$, 
and scaling from the range of [$w_{min}$, $w_{max}$] to [0, 255] with the idea that weight matrix has the same matrix form as an image channel. 
Therefore, the $\mathbf{W}_v$ are trained for two objects during the hiding phase: 1) the INR including the $\mathbf{W}_v$ should represent secret data $S$ well, and 2) the container image $\mathbf{C'}$ converted from the stack of the $\mathbf{W}_v$ should be close to the cover image $\mathbf{C}$. During the hiding phase, all other parts of the base network except for the $\mathbf{W}_v$ are fixed. 

For the first object, the INR $f(x_i)_\theta: \mathbb{R}^k \rightarrow \mathbb{R}^m$ are trained to approximate the data point $d_{i \in \mathcal{I}}$ of secret data $\mathbf{S}$ according to the input coordinate vectors $x_{i \in \mathcal{I}}$, where $\mathcal{I}$ is a pre-defined set of input index.
During the training, the $\mathbf{W}_v$ are trained with a secret data reconstruction loss $\mathcal{L}_{s}$, where
\begin{equation}
    \mathcal{L}_{s}(f_\theta) = \sum_{i \in \mathcal{I}} \Vert d_i - f_\theta(x_i) \Vert^{2}_{2}.
\end{equation}
For the second object, we constrain a stack of the variable weight matrices $\mathbf{W}_s \in \mathbb{R}^{N \times N \times 3}$ to resemble the cover image $\mathbf{C}\in \mathbb{R}^{N \times N \times 3}$.
That is, we aim to directly utilize scaled $\mathbf{W}_s$ as a container image $\mathbf{C'}$ by minimizing a loss
\begin{equation}
    \mathcal{L}_{c}(\mathbf{W}_s) = \Vert \mathbf{W}_s - (\mathbf{C}/255 \cdot (w_{max} - w_{min}) + w_{min}) \Vert ^2_2.
\end{equation}

However, we did not consider the quantization error caused by converting the weights to images in this section. Hence, it is imperative to address the quantization error during the hiding process.

\subsection{Quantization-Aware Training for Weight-to-Image Conversion}

It should be noted that converting weights stored in FP32 to an image stored in UINT8 can cause a significant perturbation, leading it to deviate from the converged state. This poses a major challenge and needs to be handled carefully during the process. To address this, we introduce simplified quantization-aware training (QAT) for weight-to-image conversion. Originally, 
QAT is a method that is widely used in network compression problems, which considers the quantization process during training. We use QAT with some modifications to solve the problem that the weights are shifted from the converged state during conversion.

We use a quantization module $f_{\theta_Q}$ with quantized variable weights $\mathbf{Q}_v$ and the same pre-defined weights, which simulates the quantization process during the training. 
For the simulation, the variable weights $\mathbf{W}_v$ are quantized to $\mathbf{Q}_v$ for every updates of $\mathbf{W}_v$ as below formula:
\begin{equation}
\begin{split}
& \mathbf{Q}_v = \textbf{ROUND}(255\cdot(\mathbf{W}_v - w_{min}) / (w_{max} - w_{min})) \\
& \mathbf{Q}_v \leftarrow \mathbf{Q}_v/255 \cdot (w_{max} - w_{min}) + w_{min}.
\end{split}
\end{equation}
The quantization module $f_{\theta_Q}$ is used to calculate the gradient $d\mathcal{L}/d\textbf{Q}_v$. 
As explained earlier, the total loss for the quantized network $\mathcal{L}$ is:
\begin{equation}
% \begin{split}
%     & \mathcal{L}_{s} = \sum_{i \in \mathcal{I}} \Vert d_i - f_{\theta_{Q}}(x_i)\Vert^{2}_{2} \\
%     & \mathcal{L}_{c} = \Vert \mathbf{Q}_h - (\mathbf{C}/255 \cdot (w_{max} - w_{min}) + w_{min}) \Vert ^2_2 \\
%     & \mathcal{L} = \mathcal{L}_{c} + \beta\cdot \mathcal{L}_{s}.
    \mathcal{L} = \mathcal{L}_{s}(f_{\theta_{Q}}) + \beta \cdot \mathcal{L}_{c}(\mathbf{Q}_{s})
% \end{split}
\end{equation}
where $\mathbf{Q}_s$ is a stack of $\mathbf{Q}_v$. To back-propagate the gradient, we convert $d\mathcal{L}/d\textbf{Q}_v$ into $d\mathcal{L}/d\textbf{W}_v$ with straight through estimator (STE) \cite{STE}.
Therefore, $\textbf{W}_v$ is finally updated as follows:
\begin{equation}
\begin{split}
    & d\mathcal{L}/d\textbf{W}_v \approx \textbf{STE}(d\mathcal{L}/d\textbf{Q}_v) \\
    & \textbf{W}_v \leftarrow \textbf{W}_v - \alpha\cdot d\mathcal{L}/d\textbf{W}_v.
\end{split}
\end{equation}
The provided Fig.~\ref{method} illustrates the proposed steganography framework as a whole.

\section{Experiments}

We experiment with the proposed method on three tasks: video-into-image steganography, audio-into-image steganography, and shape-into-image steganography. These tasks have not been explored well due to the difficulty that secret data has higher dimensions than cover data or contains temporal information. We demonstrate the performance and flexibility of our method by solving these challenging tasks for the first time.

\begin{figure}[t]
\begin{center}
\centerline{
\includegraphics[width=1\columnwidth]{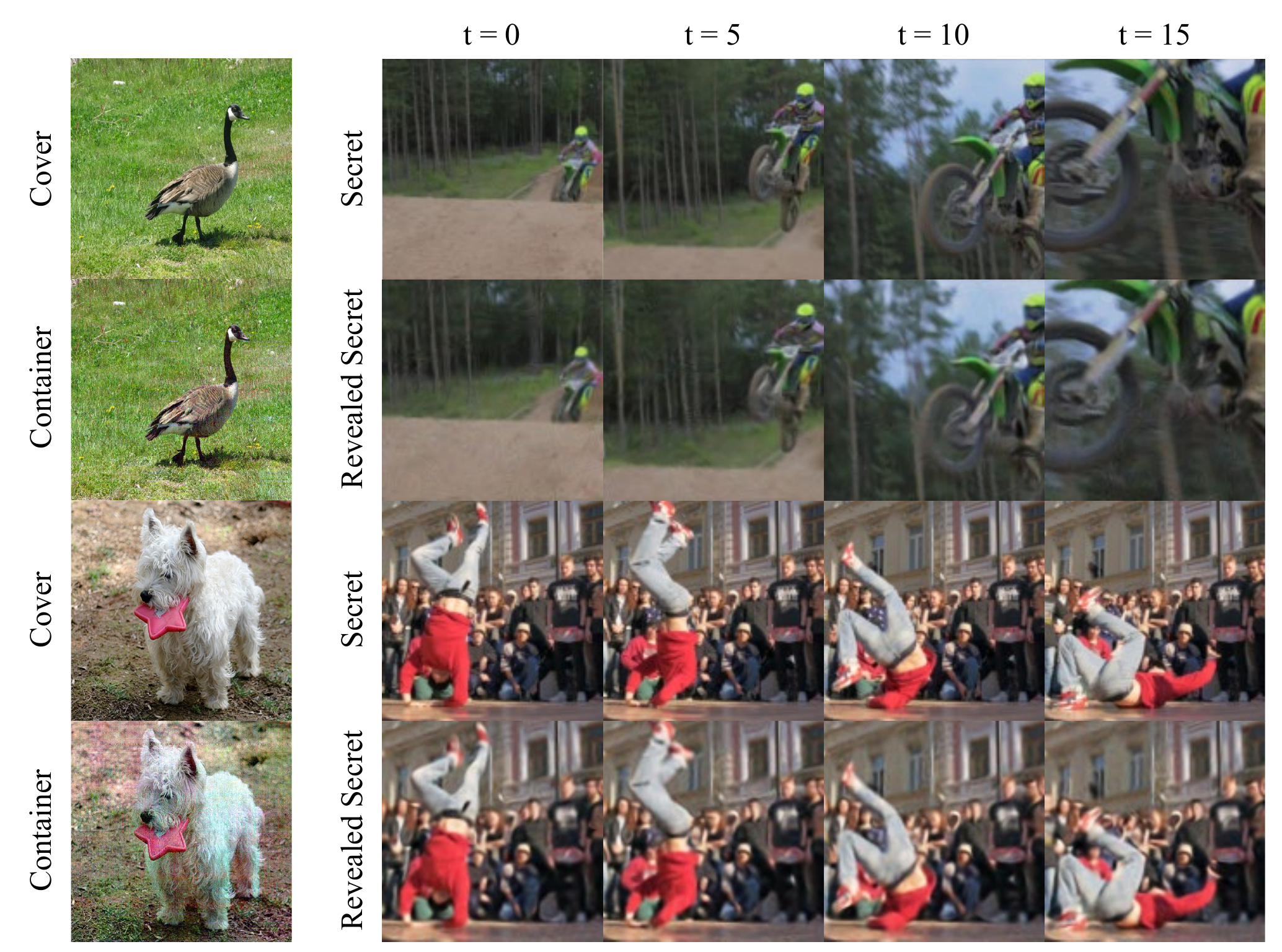}
}
\caption{Qualitative results of video-into-image steganography.}
\label{result-video}
\end{center}
\end{figure}

\subsection{Experimental Setting}

\textbf{Datasets.}
In common to all three tasks, cover images are randomly sampled from the ImageNet \cite{5206848} by the number of data in the corresponding secret dataset, and all sampled cover images are resized to 512 $\times$ 512.

For video-into-image steganography, we use the Densely Annotated VIdeo Segmentation dataset (DAVIS) \cite{Perazzi_CVPR_2016} as a secret dataset. We resize the secret videos to 128 $\times$ 128 and extract 16 frames from the videos. For audio-into-image steganography, we hide audio data from GTZAN music genre dataset \cite{tzanetakis2002manipulation, tzanetakis2002musical}. Each hidden audio data is cropped to a length of 100,000 samples with a 22,050 sample rate (4.54 seconds). For shape-into-image steganography, we selected two samples, Stanford Bunny and Dragon, as secret three-dimensional (3D) shapes from The Stanford 3D Scanning Repository \cite{stanford3d}.

\textbf{Details.}
We use SIREN \cite{NEURIPS2020_53c04118} with four hidden layers (total six layers) as the architecture of INRs to hide video and audio, and IGR \cite{pmlr-v119-gropp20a} with six hidden layers (total eight layers) as the architecture of INRs to hide shapes.  For SIREN and IGR, $\{\mathbf{W}_1,\mathbf{W}_2,\mathbf{W}_3\}$ and $\{\mathbf{W}_1,\mathbf{W}_2,\mathbf{W}_4\}$ are the variable weights and represent color channels of the container image, respectively. The remaining pre-defined weights $\mathbf{W}_p$ are fixed in an initialized state. The INRs are trained for 5,000 steps for video and shapes, and 20,000 steps for audio using Adam optimizer. 
In addition, the learning rate of the optimizer is set to $1 \times 10^{-3}$ for video and audio and set to $1 \times 10^{-4}$ for shapes.

\subsection{Experimental Results}

\textbf{Video-into-image steganography.}
The Average Pixel Discrepancy (APD) of the cover and secret is calculated as the $L_1$ distance between the cover and container and that between the frames of the secret video and revealed secret video, respectively. In addition to APD, we report Peak Signal-to-Noise Ratio (PSNR), Structural Similarity (SSIM), and Perceptual Similarity (LPIPS) in the same way in Table~\ref{result-table-video}. The qualitative results for the video-into-image steganography can be confirmed in Fig.~\ref{result-video}.

\begin{table}[h]
\centering
\begin{small}
% \resizebox{\columnwidth}{!}{
\begin{tabular}{c|cccc}
  \toprule
Errors & APD $\downarrow$ & PSNR $\uparrow$ & SSIM $\uparrow$ & LPIPS $\downarrow$ \\  
\midrule
Cover & 20.32 & 19.77 & 0.5191 & 0.5262  \\ 
Secret & 5.79 & 29.98 & 0.9263 & 0.1333  \\ 
\bottomrule
\end{tabular}
% }
\end{small}
\caption{Performance of the video-into-image steganography.}
\label{result-table-video}
\end{table} 

\begin{figure}[t]
\begin{center}
\centerline{
\includegraphics[width=1\columnwidth]{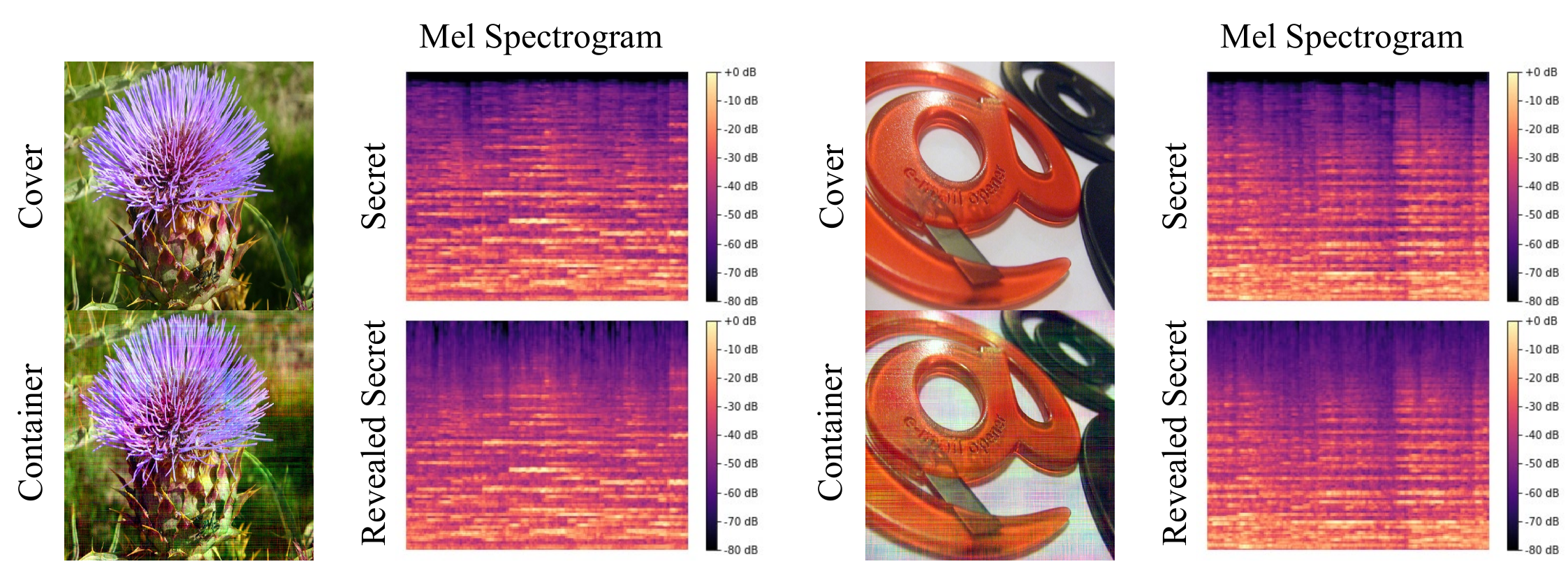}
}
\caption{Qualitative results of audio-into-image steganography.}
\label{result-audio}
\end{center}
\end{figure}

\textbf{Audio-into-image steganography.}
For evaluation of reconstructed audio, we report Absolute Error (AE) and Signal-to-Noise Ratio (SNR) between the secret and revealed secret instead of APD and PSNR in Table~\ref{result-table-audio}. In addition, the audios of the secret and revealed secret are visualized as mel-spectrograms in Fig.~\ref{result-audio}.

\begin{table}[h]
\centering
\begin{small}
% \resizebox{\columnwidth}{!}{
\begin{tabular}{c|cccc}
  \toprule
Errors & APD/AE $\downarrow$ & PSNR/SNR $\uparrow$ & SSIM $\uparrow$ & LPIPS $\downarrow$ \\  
\midrule
Cover & 13.73 & 23.33 & 0.7139 & 0.3948  \\ 
Secret & 2.84 & 17.92 & 0.9328 & -  \\ 
\bottomrule
\end{tabular}
% }
\end{small}
\caption{Performance of the audio-into-image steganography.}
\label{result-table-audio}
\end{table} 

\begin{figure}[t]
\begin{center}
\vskip -25pt
\centerline{
\includegraphics[width=1\columnwidth]{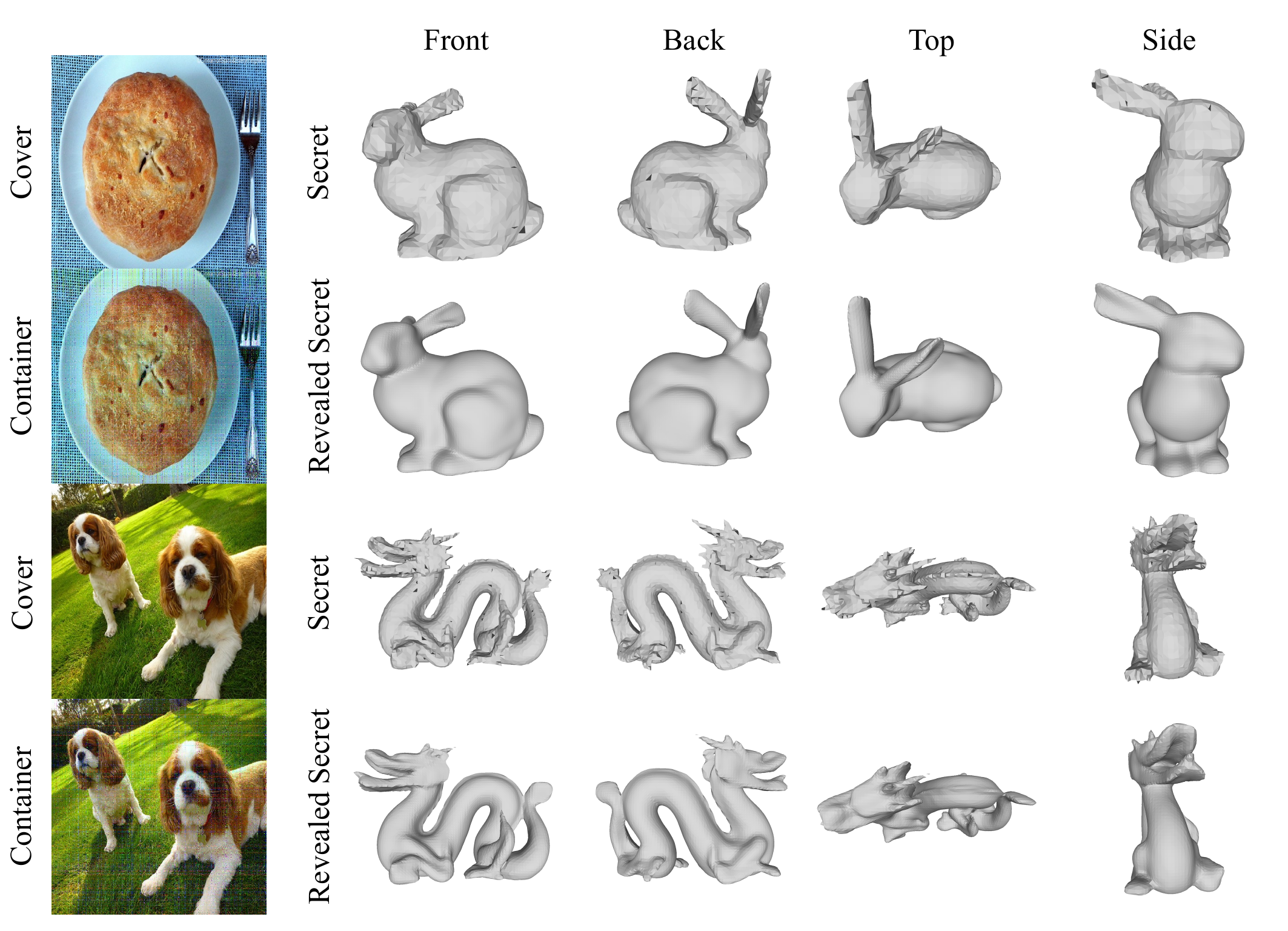}}
\caption{Qualitative results of shape-into-image steganography.}
\vskip -25pt
\label{result-shape}
\end{center}
\end{figure}

\textbf{Shape-into-image steganography.} 
We experiment with the proposed method on the shape-into-image steganography to demonstrate that the proposed method has a significantly broader scope of application than conventional deep steganography methods, as 3D shapes are fundamentally different from the data that previous deep steganography methods attempted to conceal.
In Fig.~\ref{result-shape}, we present the reconstructed Stanford Bunny and Dragon from various viewpoints.

\begin{figure}[t]
\begin{center}
\centerline{
\includegraphics[width=0.8\columnwidth]{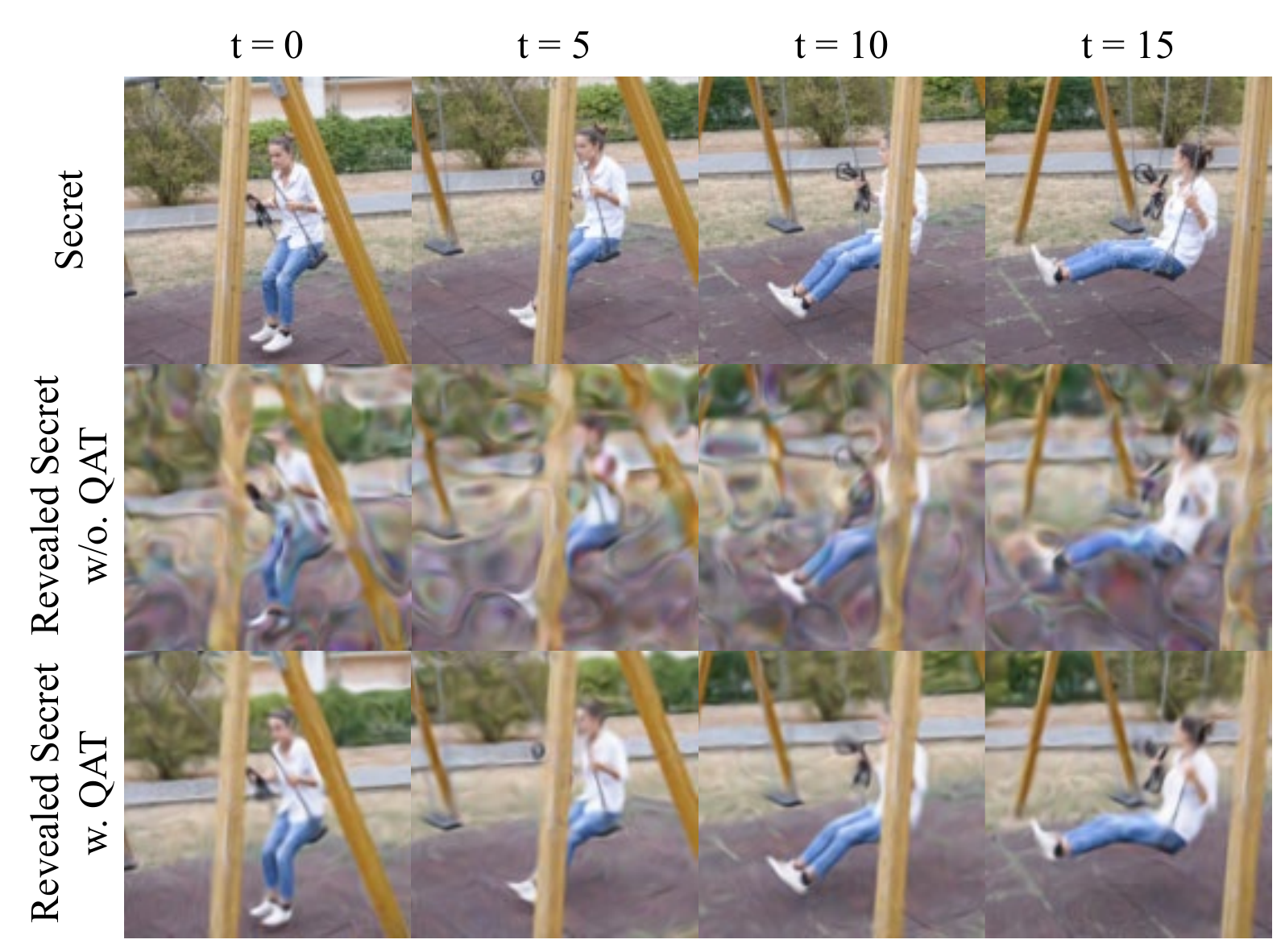}
}
\caption{Effectiveness of quantization-aware training.}
\vskip -25pt
\label{result-ablation}
\end{center}
\end{figure}

\textbf{Ablation study.}
We measure the performance difference of video-into-image steganography depending on whether QAT is applied or not. 
The experimental results in Table~\ref{result-table-ablation} show that when QAT is not applied, the revealed secret is very altered from the original due to the weight shifts by quantization. We also present the visual effect of QAT through the sample included in Fig.~\ref{result-ablation}.

\begin{table}[h]
\centering
\begin{small}
\begin{tabular}{c|cccc}
\toprule
Errors & APD $\downarrow$ & PSNR $\uparrow$ & SSIM $\uparrow$ & LPIPS $\downarrow$ \\  
\midrule
Cover w/o. QAT & 19.17 & 20.37 & 0.5700 & 0.4785  \\ 
Secret w/o. QAT& 27.76 & 17.73 & 0.5094 & 0.4279  \\
Cover w. QAT & 20.32 & 19.77 & 0.5191 & 0.5262  \\ 
Secret w. QAT & 5.79 & 29.98 & 0.9263 & 0.1333  \\ 
\bottomrule
\end{tabular}
\end{small}
\caption{Ablation study results on QAT.}
\vskip -10pt
\label{result-table-ablation}
\end{table} 

\subsection{Discussion and Future Work} Through our experiments, we show that it is possible to hide secret data of various modalities with only subtle changes to the cover image. 
However, there is a limitation that the format of cover data is fixed as an image, and there is some loss in high-frequency information for revealed secret. Since this paper introduces a steganography solution using INRs for the first time, we believe that there is still room for improvement and various applications.
We also look forward to future work to improve the robustness of the proposed approach against container distortions such as blur or JPEG compression.

\section{Conclusion}
We proposed a novel approach to deep cross-modal steganography utilizing INRs, which enables the hiding of data of diverse modalities within images. We believe that our proposed method will stimulate further investigation, as it introduces a fresh research direction for deep steganography.

\section{Acknowledgement} 
This work was supported by Artificial intelligence industrial convergence cluster development project funded by the Ministry of Science and ICT(MSIT, Korea)\&Gwangju Metropolitan City.

\bibliographystyle{IEEEbib}
\bibliography{references}

\end{document}